\documentclass[runningheads,a4paper]{llncs}

\usepackage{amssymb}
\usepackage{graphicx}

\usepackage{url}
\urldef{\mailsa}\path|{ingrid.dillo, rene.van.horik, andrea.scharnhorst}@dans.knaw.nl|    
\newcommand{\keywords}[1]{\par\addvspace\baselineskip
\noindent\keywordname\enspace\ignorespaces#1}

\begin{document}

\mainmatter  

\title{Training in Data Curation as Service in a Federated Data Infrastructure - the \emph{FrontOffice--BackOffice Model}}

\titlerunning{Training in Data Curation}

\author{Ingrid Dillo \and Rene van Horik \and Andrea Scharnhorst}
\authorrunning{Dillo}

\institute{Data Archiving and Networked Services, Anna van Saksenlaan 10, \\
2593 HT The Hague , The Netherlands
\mailsa\\
\url{http://www.dans.knaw.nl}}

\toctitle{Training in Data Curation as Service in a Federated Data Infrastructure - the \emph{FrontOffice--BackOffice model}}
\tocauthor{Dillo}
\maketitle

\begin{abstract}
The increasing volume and importance of research data leads to the emergence of research data infrastructures in which data management plays an important role. As a consequence, practices at digital archives and libraries change. In this paper, we focus on a possible alliance between archives and libraries around training activities in data curation. We introduce a so-called \emph{FrontOffice--BackOffice model} and discuss experiences of its implementation in the Netherlands. In this model, an efficient division of tasks relies on a distributed infrastructure in which research institutions (i.e., universities) use centralized storage and data curation services provided by national research data archives. The training activities are aimed at information professionals working at those research institutions, for instance as digital librarians. We describe our experiences with the course \emph{DataIntelligence4Librarians}. Eventually, we reflect about the international dimension of education and training around data curation and stewardship.   
\keywords{data curation, data management, training, data sharing, data archive, digital libraries, education, science policy, documentation}
\end{abstract}
\section{Introduction}
A research archive can be depicted as a safe haven for research data, carefully selected, documented and stored for future consultation. Accordingly, the core tasks of a data archivist could be imagined to be confined to proper documentation, and the care for material preservation. In short: "Our service starts where others drop the data"\footnote{Personal communication Henk Koning, former Technical Archivist at DANS}. The current practices of archivists seem to deviate from such an archetype to a large extent. This \emph{turn of tables} can best be understood by a recall to the history of archival sciences.
\begin{figure}[h]
\centering
\includegraphics[height=9cm]{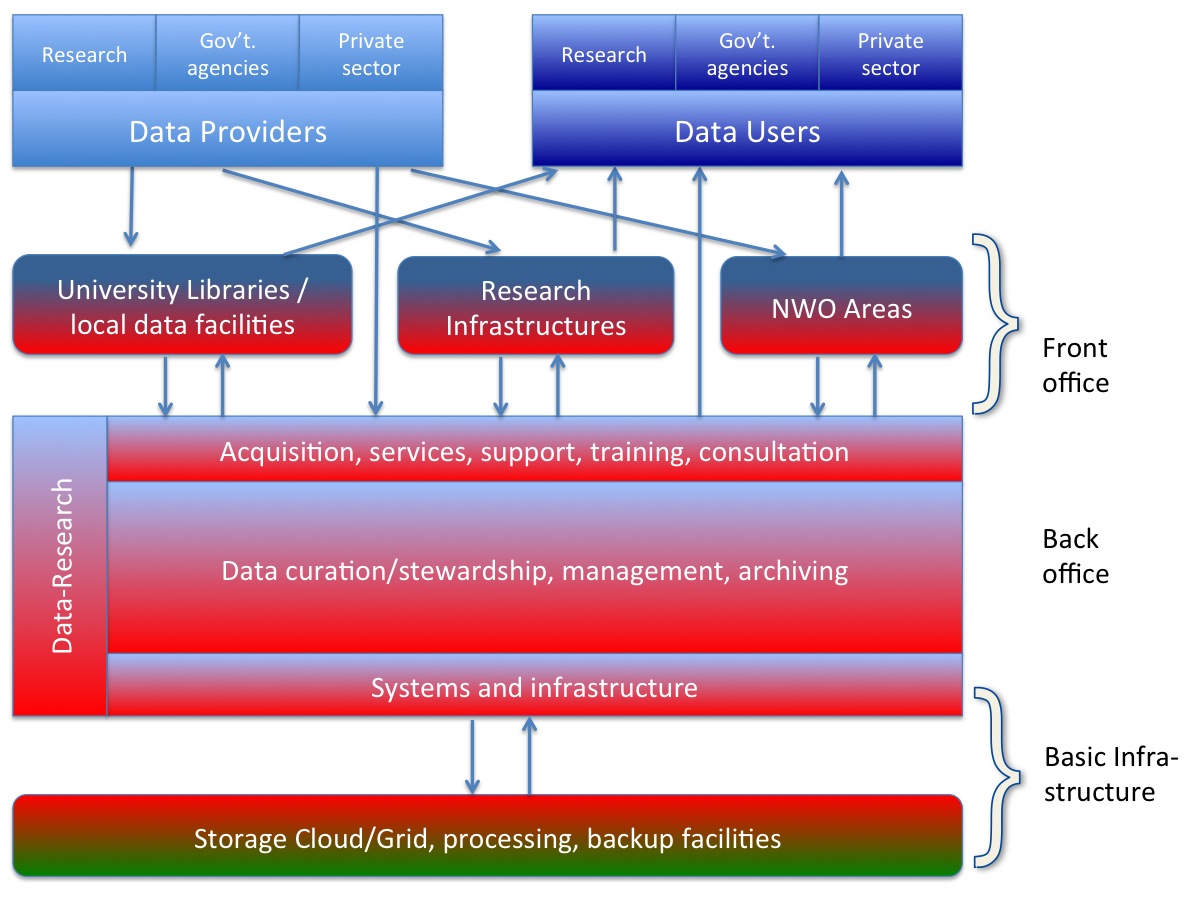}
\caption{The federated data infrastructure - a collaborative framework. Scheme designed by Peter Doorn based on the \emph{Collaborative Data Infrastructure} as envisioned in \cite[p. 31]{hlg2010riding}}.
\label{fig:datainfra}
\end{figure}
In general, for archives of research data the same principles hold as for any
other archive. In 1898, in the \emph{handbook}, one of the foundational texts
in archival sciences \cite{handleiding1898}, Muller, Feith, and Fruin describe
the archive as an organic entirety whose function cannot be determined \emph{a
priori}. On the contrary, its function needs to be defined and redefined
depending on the development of the institution (i.e., a board or government)
whose selected traces it is obliged to archive. In other words, Muller et
al. describe a \emph{co-evolution} of the institution and its archive. This
view applied to a research data archive, the corresponding institution is none
other than the science system. From out this viewpoint, it is not surprising
that the profound changes in scientific practice \cite{wouters2012virtual} and
scholarly communication \cite{borgman2007scholarship} influence the
expectations placed on a data archive or, more specifically, a sustainable
digital archive (Trusted Digital Repository). The changing modes of scholarly
communication and practice alter the form and content of what is seen worth to
be preserved.  \cite{doorn2007introduction} Changing research practices
require new negotiations on the division of labor. Who is responsible for
setting up digital research infrastructures including virtual research
environments - the information service providers such as Trusted Digital Repositories (TDRs) or the research institutions? Who takes care of the preparation of (meta)--data and formats prior to archiving? Who should preserve software tools - the labs which developed them or the archive together with 'data' for which they have been developed? \\
The high volatility of the environment in which archives are currently operating influences their function as reliable, stable reference point for important information. Open Access, Data Management Plan, Data Stewardship, Data Curation, Trusted Digital Repositories, BigData and SmartData are some of the floating around buzzwords of the last decade. They stand for the struggle to identify and communicate most urgent trends and to coordinate actions across the different stakeholders in the field of data curation. Important to note here is the reference model for Open Archival Information Systems (in short OAIS model, ISO 14721:2012), a model foundational for the discussion of structure and function of any archive. Its key elements are Ingest, Archival Storage, Data Management, Administration, Preservation Planning and Access. Allison emphasizes that the OAIS model is not an architectural model for implementation, but instead offers a shared terminology.  \cite{allison2007evaluation}.  
Inside of our own organization, Data Archiving and Networked Services (DANS), the OAIS model is often used in discussions about \emph{internal} workflows and their improvement and further development. In this paper, we focus on institutional networks \emph{around} an archive as DANS. Hereby we rely on schemata as depicted in Fig.~\ref{fig:datainfra} which sketch the complexity of the
\emph{research data landscape}, its stakeholders and infrastructure
\cite{hlg2010riding}. Coming back to it later, in a first step we can use this
scheme in an exercise to \emph{locate} a TDR such as 
DANS. Starting at the bottom of Fig.~\ref{fig:datainfra} the basic
(technical) infrastructure entails storage. In the Netherlands this level of Basic Infrastructure is provided by 
SURFsara, the Dutch network of
computing facilities whose services DANS is using itself. The following three
levels could be seen as the heart of activities of an archive of
\emph{digital} research data. They form a kind of back-office. The three  boxes at the next level, labeled as
front office, contain the funding agencies, as NWO\footnote{\url{www.nwo.nl}} in the Netherlands,
university libraries, and research infrastructures such as
CLARIN\footnote{\url{www.clarin.eu}}, or DARIAH\footnote{\url{www.dariah.eu}},
which are in themselves complex organizations. They could be seen as 'clients'
of an archive. But actually, DANS is also part of them. The same holds true for the top level of data providers and users. DANS as part of research infrastructures harvests information from other data providers. With its own research and development activities it is even part of the data production cycle. In short, DANS plays different roles in different contexts and, therefore, can be located at many places in this scheme. Correspondingly, at DANS a variety of different activities take place. In the next section, we discuss how, together with this increase in complexity, the need emerges to build alliances and to coordinate actions among different institutional players in the \emph{data landscape}. At the core of the paper we propose a specific model to articulate possibilities of collaboration, coordination, and division of labour. We report about steps towards its concrete implementation at the Dutch national level. At the end of the paper we discuss links to international developments. 
\section{The archivist as a consultant}
DANS is one of the national research data archives in the Netherlands. With roots in the social sciences and humanities back to the 1960s, in its current form, it was founded in 2005 as an institute of NWO - the Netherlands Organization for Scientific Research and the KNAW - the Royal Netherlands Academy of Arts and Sciences. DANS is primarily an information service institute and, despite of a small in-house research group, not a research institute. This makes DANS much more comparable to a classical, stand-alone archive. \\
The mission of DANS it to promote sustained access to digital research
data. For this purpose, DANS encourages researchers to archive and reuse data
in a sustained manner, e.g. through the online (self)archiving system
EASY\footnote{\url{www.easy.dans.knaw.nl}}. DANS also provides access, via
NARCIS.nl\footnote{\url{www.narcis.nl}}, to thousands of scientific datasets,
e-publications and other research information in the Netherlands. EASY and
NARCIS are two services which form the core of DANS. In difference to many
other knowledge-domain specific archives, DANS operates cross-disciplinary
with a focus on social sciences and humanities. It is also an exclusively
digital archive and it is placed - as an institution - outside the Dutch
university system. All this together positions DANS as a gateway to the
diverse Dutch research data landscape and as a hub in it. Activities and practices at DANS can be ordered along three dimensions:
\begin{itemize}
  \item \emph{Archive}: selection, preservation, and description of data collections
  \item \emph{Research and Development}: maintenance and development of the ICT infrastructure for seamless access and exploitation and for long-term preservation
  \item \emph{Science Policy}: influence on research data policies and data
  curation strategies on the national and international levels 
\end{itemize}
The first dimension corresponds to a large extent to the image of a
traditional research archive. But due to ongoing ICT innovations both in the
area of research as well as of information services, a digital archive cannot
operate without means to adopt its technological backbone to those
innovations. The process of adopting and inventing services entails to a large
extent what Andrew Prescott called "tinkering", when he compared practices at
digital libraries with the craftsmanship needed in labs and workshops in the
high-time of industrialization \cite{prescott2012sheffield}. ICT is usually
depicted as an efficiency engine. What is often forgotten is the existence of
a transition period during which old and new forms of practices coexist. On
the work floor, this means that traditional services of acquisition, community
support, and documentation are pursued in parallel to designing new workflows,
testing and implementing them. So, before ICT leads to more efficiency,
temporarily the actual workload often increases. Project-based work and
external funding for projects can only partly buffer this extension of
activities at an archive. On top of archiving and related R\&D, the changing
environment in which the archive operates requires continuous
attention. Hence, a third dimension - science policy - appears. Participation in national and international networks of research infrastructures require substantive investment of time. \\
The point we make is that the current portfolio of activities at information
service institutions is much more diverse than in the past. For DANS this
changing role of \emph{an archive} is reflected in its name as \emph{Data
Archiving and Networked Services}. Among the increased portfolio of
activities, \emph{consultancy} plays a special role  \cite{dans2010plan}. It
appears in many forms: in the foundation of a Data Seal of Approval for
TDRs\footnote{\url{www.datasealofapproval.org}}, in the advisory role in
research projects, in contributions to data policy documents, and in training
activities. Consultancy contributes to knowledge diffusion around data
curation practices and the coordination of data management at a national (partly also international) level. It also supports the emergence of a distributed network structure which we describe in the next section.
\section{Strategic alliance between archives and libraries - the \emph{FrontOffice -- BackOffice model}}
Profound and timely data management together with a sustainable storage of
data -- during and after the research -- are indispensable preconditions for
sharing data. It is of great importance that universities and other research
institutions develop a clear data policy themselves. An adequate infrastructure is needed to coordinate and implement those policies. In the Netherlands, with its rich institutional landscape of information service providers and research institutions, we encounter a discussion around a federated data infrastructure. It is quite clear that no single organization will be able to deliver individually tailored support for all possible data depositors. It it also clear that it is not possible for a single organization to provide services across all levels, from storage up to interactions with individual researchers. In order to create a sustainable national infrastructure for data management and curation, it is important to support a network of \emph{local} data stewards close to the actual scientific practice combined with centralized services. Fig.~\ref{fig:datainfra} designs such a federated data infrastructure. It introduces at the same time a \emph{FrontOffice--BackOffice model} (FO--BO model) as part of it. 
\subsection{Description of the model}
\begin{figure}[h]
\centering
\includegraphics[height=9cm]{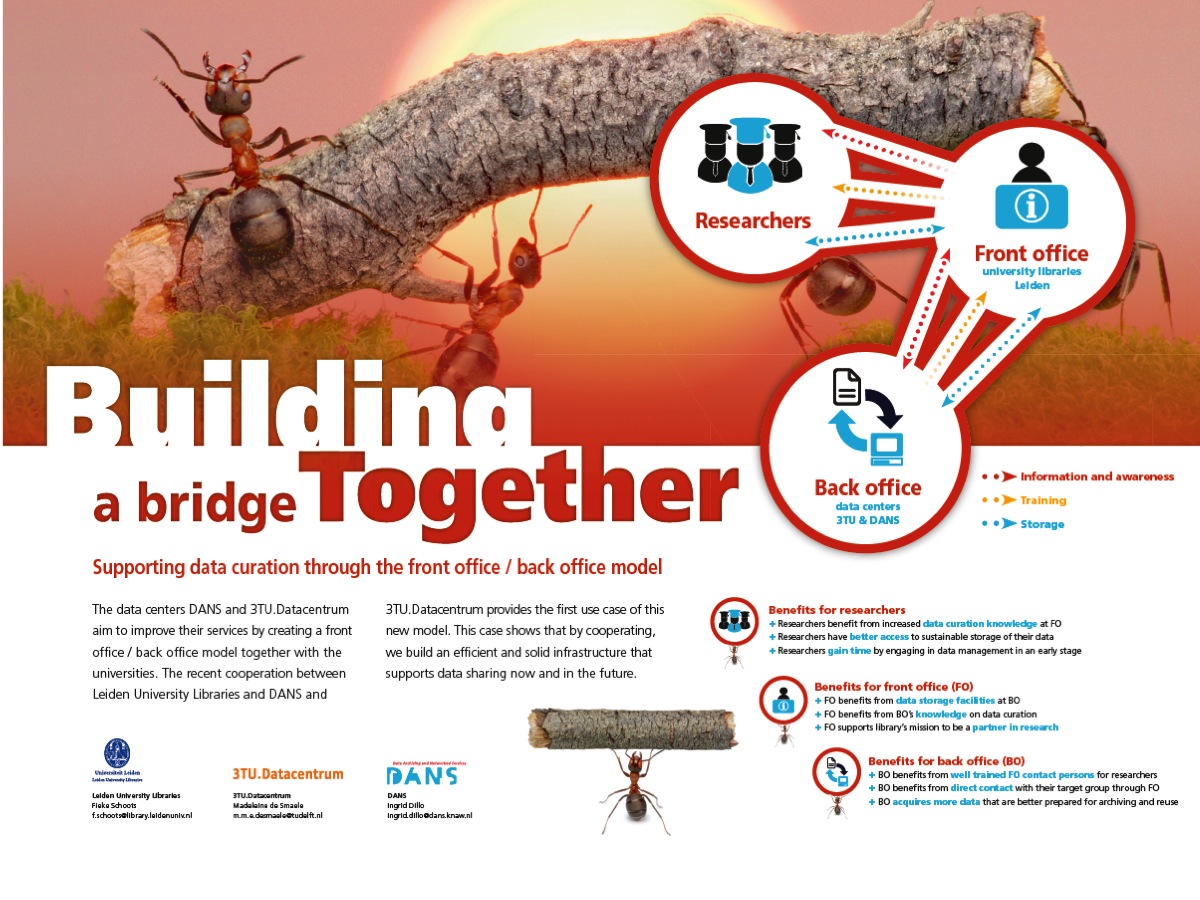}
\caption{Poster about the FrontOffice--BackOffice model. Designed by Carolien van Zuilekom, Fieke Schoots, Madeleine de Smaele and Ingrid Dillo}
\label{fig:foboposter}
\end{figure}

The FO--BO model clarifies the interaction between researcher and information
service provider concerning research data management. It also clarifies the
relation among different information service providers. Front offices should
be placed at institutions where research takes place in order to support the
research community at those institutions. An example could be a front office
as part of a university library. The front office is responsible for raising
awareness for data sharing and re-use, for taking care of the local data
management, and for organizing training for researchers. Virtual Research
Environments (VRE's) could be also part of the service at a front office. In
particular, temporary data archiving on platforms as Sharepoint or Dataverse
could be part of the VRE's. Once a research project is finished the front
office - in consultation with the back office - takes care of the transfer of
data to a TDR. So, data acquisition is an inherent part of the front office tasks. \\
The core tasks of the back office consist in the storage and documentation of
research data which arrive via the front offices. The back office provides
access to data, and possibly enriches and links data. The back office acquires
expert knowledge around data management, and the long-term, sustainable and
persistent archiving of research data. Part of the back office portfolio is to
disseminate this expertise by means of training of information professionals,
such as data librarians/managers/stewards, working at front offices. The back
office acts as an expertise centrum for the front office and as an innovation centrum concerning new trends in data curation. Fig.~\ref{fig:foboposter} summarizes the benefits of the model for researchers, front offices and back office organizations. By means of the FO--BO model we also try to reduce the complexity of interactions in the data infrastructure. With this model the role of DANS (and of comparable institutions) is restricted to the back office function. In the next subsection we report about one key element of the model: training for front office personal.
\subsection{Implementation - the \emph{DataIntelligence4Librarians}} 
In the FO--BO model training for information professionals is part of the back office portfolio. The \emph{DataIntelligence4Librarians} course is an example for such a training. Organized by the 3TU.Datacenter\footnote{The 3TU.Datacenter -- a network organization of the university libraries of Delft University of Technology, Eindhoven University of Technology, and the University of Twente -- offers facilities for the preservation and the sustained availability of technical research data, similar to the services at DANS (see \url{http://data.3tu.nl/repository/})} and DANS, it is based on an earlier course of the 3TU.Datacenter developed for data-librarians. The currently envisioned audience reaches from staff at libraries to everybody interested in the topic independently of the disciplinary background. 
\begin{figure}[h]
\centering
\includegraphics[height=9cm]{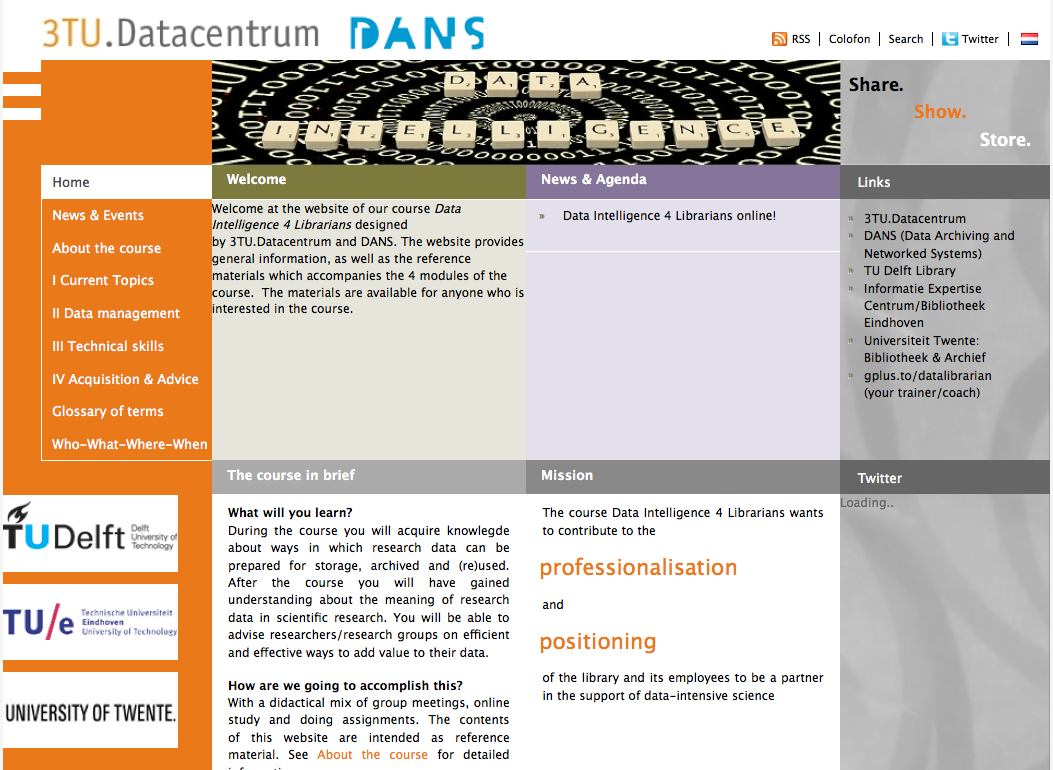}
\caption{Snapshot of the website http://dataintelligence.3tu.nl/en/home/ - host of the course "DataIntelligence4Librarians"}
\label{fig:dataintelligence}
\end{figure}
\subsubsection{Description of the course}
The course design fits into the professional education format. It combines
distance learning with four face2face (f2f) sessions and maintains next to an
eLearning environment also a public website (see
Fig.~\ref{fig:dataintelligence}, in Dutch) with background material. Google
Plus was used as the platform for the eLearning part. Participants are supposed to study theoretical parts as homework. Between the f2f sessions more homework is assigned. The website contains a description for the first practical task. More of them are distributed in the eLearning environment. Both coach and participants give feedback at f2f sessions as well as on-line. Didactically, feedback and knowledge sharing is used as an important element next to knowledge transfer. \\
During the first f2f session, an introduction into the course and the
eLearning environment is given. An introduction into the module \emph{Data
Management} follows and homework is assigned labeled \emph{State-of-Art
Map}. This task starts with reading a report, and continues with a number of
search tasks using the phrase \emph{research data management} across
bibliographic databases (Scopus, Web of Science), but also in Twitter and
Google. Participants are advised to subscribe to specific mailing lists to get
an impression of the actual discussion around the topic. At the second f2f
session participants presents their resulting map. The module \emph{Technical
Skills} is introduced and tools (3TU.Databrowser, DANS/EASY) are
demonstrated. The third f2f meeting starts with the same scheme of sharing
homework and getting feedback. Content-wise the module for this session is
\emph{Acquisition and Consultation Skills}. Specific attention is given to the
question how to overcome barriers for data sharing. The instrument of a
\emph{Data Interview} with possible data depositors is introduced. During the
fourth and last session the acquisition assignment is discussed and the course
is evaluated. At the end of the course a certificate is issued under the
condition that all sessions have been attended and the tasks have been fulfilled. During the modules, different experts from the organizing institutions give guest lectures. Examples of topics are legal aspects, issues of data selection, audit and certification of TDRs, and the FO--BO model itself.
\subsubsection{Experiences with the course}
One goal of the course is to sharpen the insight into the role of research data in scientific practices. Eventually, the participant should be able to advise and inform researchers how data curation can enhance data use and re-use. In summary, the goals are rather diverse and broad compared with the limited time of the course.\\
So far, the course has been run three times: February 2012 to June 2012 (16
participants, organized by 3TU.Datacentre), September 2012 to December 2012
(16 participants), and  February 2013 to May 2013 (13 participants). The last
two events have been organized in collaboration between the 3TU.Datacentre and
DANS. Most of the participants were information professionals, either working
at a library or archive, or for one of the network organizations, such as
SURF. In the evaluation, the participants named a couple of critical points
not unusual for distance learning. Among them are problems with the eLearning
environment, or the spreading out of the course over a rather long
period. Another critical remark concerns the demonstrations. Obviously the
participants did not seek hands-on experiences with a tool, platform, or
interface. They seemed to be more interested in guidance and factual
information in the area of data curation. This springs also out from the
positive reactions. Information about actual developments from experts
involved in the practice of data curation have been highly
appreciated. Further, a need to get to know each other and to learn from each
others practices is articulated. This holds true even for a small country as the Netherlands. One of the suggestions of the participants was to form a \emph{special interest group}. 
\section{Conclusions}
In this paper we discussed changing portfolios of responsibilities for
archives and libraries. Data infrastructures emerge in response to data
science, open access, and data sharing policies. In the making of a data
infrastructure, the division of tasks between different information service providers needs to be re-negotiated. We present a federal data infrastructure with a layered architecture including a \emph{FrontOffice--BackOffice model}. This model allows to articulate different roles in the interaction with research communities, the acquisition of expert knowledge, and the provision of data management services. The model is in line with the \emph{Data pyramid} \cite{hlg2010riding} which classifies data according to permanence and function. Data management is tailored towards certain classes of data and specialization in data curation is allocated to different organizations. Front offices, naturally to be placed at academic libraries, take care of data management for \emph{transient and cyclic data} produced by individuals and research communities. Trusted Digital Repositories as DANS act as back office and take care for \emph{patrimonial data}. They also become expertise center and knowledge transfer hubs for data curation. \\ 
Training plays a key role in the FO--BO model. It is a way to disseminate the
idea of the model. At the same time, it is an instantiation of the model.  The
experiences in the Netherlands are encouraging. Several Dutch universities
signaled interest in this approach and the challenge is now to implement more front offices there. At the same time, a coordination among possible back office organizations is needed. DANS recently signed an coalition agreement with the 3TU.Datacenter to cooperate more closely and to foster the FO--BO model. This coalition, \emph{Research Data Netherlands}, is open to any other Dutch TDR with at least a Data Seal of Approval. To shape the role of back offices as centers of expertise and innovation is another way to make the model attractive and reliable. To give an example, there is a growing need for auto-ingest of larger data collections. Another shared issue is the question of a sustainable cost model for data archiving. Exploration of these issues needs to be done locally and shared in collaboration. \\ \\
Returning to the issue of training, as we argue in this paper, in the short
run, there is an urgent need for education among information professionals. In
the mid term,  these efforts could be connected to comparable modules in
curricula for future information professionals, e.g. at iSchools. The FO--BO
models contains training also as a part of front office activities. This is in line with efforts in the framework of digital librarianship to develop  modules for information literacy and data stewardship at many universities. The APARSEN project that aims at establishing a virtual centre of excellence on digital preservation carried out a survey concerning the European training landscape in this area \cite{aparsen2012survey}. The DataIntelligence4Librarians course fits very well to outcomes of this survey. A coordination between those different training activities will support further professionalization. Shared textbooks, syllabi, best practices guidelines could also help to keep locally provided on-line material up to date. \\
Our experiences show that a natural alliance between (digital)
archives and libraries exist which is worth to be explored in daily
practice. Current science policies emphasizes the role of data and their
re-use. The envisioned coupling of funding with data-sharing and archiving,
the Linked Open Data movement, and the rise of data science will put more
pressure on information service institutions, but at the same time also offers
new possibilities. To form alliances and to coordinate actions seems to be the only possible answer. Here, libraries and archives are natural partners because as Ross formulated "when we reflect on the core of digital libraries we easily observe that they may be libraries by name, but they are archives by nature" \cite{ross2012digital}.

\section{Acknowledgement}
The following colleagues have been involved in the DataIntelligence4Librarians course. Nicole Potters, Marina Noordegraaf, Madeleine de Smaele, Ellen Verbakel (from the 3TU.Datacenter) and Rene van Horik, Caroline van Zuilekom, Marion Wittenberg, Ingrid Dillo (from DANS). 

\begin{thebibliography}{4}

\bibitem{handleiding1898} Muller, S., Feith, J.A., Fruin, R.: Handleiding voor het Ordenen en Beschrijven van Archiven. Erven B. Van Der Kamp. Groningen. 1920. 2ed. Reprinted in: Horsman, P.J., Ketelaar, F.C.J., Thomassen, T.H.P.M.: Tekst en Context van de Handleiding voor het Ordenen en Beschrijven van Archiven van 1898. Verloren, Hilversum 1998.

\bibitem{wouters2012virtual} Wouters, P., Beaulieu, A., Scharnhorst, A., Wyatt, S.: Virtual Knowledge: Experimenting in the Humanities and the Social Sciences. MIT, Cambridge, Mass. 2012.

\bibitem{borgman2007scholarship} Borgman, C.: Scholarship in the digital age: Information, infrastructure, and the Internet. MIT, Cambridge, Mass. 2007.

\bibitem{allison2007evaluation} Allison, J.: OAIS as a reference model for repositories. An evaluation. Report UKOLN University of Bath, 2007 \url{http://eprints.whiterose.ac.uk/id/eprint/3464}.

\bibitem{doorn2007introduction} Doorn, P., Tjalsma, H.: Introduction: archiving research data. Archival Science 7(1), 1--20 (2007). DOI 10.1007/s10502-007-9054-6

\bibitem{hlg2010riding} Anonymous: Riding the wave. How Europe can gain from the riding tide of scientific data. Final report of the High Level Expert Group on Scientific Data. A submission of the European Commission. October 2010 \url{http://cordis.europa.eu/fp7/ict/e-infrastructure/docs/hlg-sdi-report}

\bibitem{prescott2012sheffield} Prescott, A.: Made In Sheffield: Industrial Perspectives on the Digital Humanities. Keynote at the Digital Humanities Congress at the University of Sheffield, 6 September 2012. (The text of this keynote lecture can be found at Andrew Prescott's blog \url{http://digitalriffs.blogspot.co.uk/2012/09/made-in-sheffield-industrial.html})

\bibitem{dans2010plan} Anonymous. Duurzame toegang tot digitale onderzoeksgegevens. Strategienota DANS (in Dutch). DANS, The Hague 2010. \url{http://www.dans.knaw.nl/sites/default/files/file/Uitgaven/Strategie/DANS STRATEGIENOTA compleet_DEF.pdf}. A summary in English with the title: Sustained access to digital research data can be found at \url{http://www.dans.knaw.nl/sites/default/files/file/jaarverslagen en strategienota/Samenvatting strategienota_UK_DEF.pdf}

\bibitem{ross2012digital} Ross, S.: Digital Preservation, Archival Science and Methodological Foundations for Digital Libraries. New Review of Information Networking 17(1), 43--68 (2012). DOI 10.1080/13614576.2012.679446

\bibitem{aparsen2012survey}Anonymous. APARSEN report: D43.1 Survey for the assessment of training material. Assessment of digital curation requirements.  Available at: \url{http://www.alliancepermanentaccess.org/wp-content/uploads/downloads/2012/12/APARSEN-REP-D43_1-01-4_1.pdf}
 
\end{thebibliography}
\end{document}